\documentclass[12pt]{article}

\usepackage{hyperref}
\usepackage{amsmath,amsfonts,amssymb}
\hypersetup{dvips,dvipdfm,colorlinks=true,urlcolor=magenta,filecolor=magenta,linktoc=page,citecolor=red,linkcolor=blue,bookmarks=true}
\usepackage{array}
\usepackage{graphicx,epstopdf}
\usepackage{multirow}
\usepackage{xcolor}
\date{}
\begin{document}
	\title{On the consequences of Raychaudhuri equation in Kantowski-Sachs space-time}
	\author{Madhukrishna Chakraborty \footnote{chakmadhu1997@gmail.com}~~and~~
		Subenoy Chakraborty\footnote{schakraborty.math@gmail.com (corresponding author)}
		\\Department of Mathematics, Jadavpur University, Kolkata - 700032, India}
	\maketitle
	\begin{abstract}
	The paper aims to study the geometry and physics of the Raychaudhuri equation (RE) in the background of a homogeneous and anisotropic space-time described by Kantowski-Sachs (KS) metric. Role of anisotropy/shear in the context of convergence and possible avoidance of singularity has been analyzed subject to a physically motivated constraint. Moreover, using a suitable transformation the first order RE has been converted to a second order differential equation analogous to a Harmonic Oscillator and criterion for convergence has been shown to be associated with the time varying frequency of the Oscillator. Finally, the paper points out a geometric and physical notion of anisotropy along with their corresponding behavior towards convergence.
	\end{abstract}
	
	\section{Introduction}
	Einstein's General theory of relativity (GR) \cite{Wald:1984rg}, \cite{Weinberg:1972kfs} is the backbone for standard model of cosmology and at the same time emerges as the most successful theory of gravity to describe physical reality. Its popularity was enhanced after the detection of gravitational waves \cite{LIGOScientific:2017vwq}, \cite{LIGOScientific:2016aoc}. Inspite of its great success, the appearance of singularity in a classical space-time is inevitable in this theory as its major drawback. This is demonstrated by the seminal singularity theorems of Hawking and Penrose \cite{Hawking:1973uf}-\cite{Hawking:1970zqf}. In this context it is worthy to mention that Raychaudhuri equation (RE) \cite{Raychaudhuri:1953yv}-\cite{Dadhich:2005qr} is the key ingredient behind the proof of these singularity theorems via the notion of geodesic focusing \cite{Borde:1987qr}, \cite{Albareti:2012se}. According to the Focusing theorem (FT) \cite{Chakraborty:2023lav}, an initially converging congruence of time-like geodesic will focus within a finite value of proper time ($\tau$) provided the strong energy condition (SEC) is satisfied by the matter content of a space-time. FT is the turning point of Einstein gravity and therefore regarded as the most vital consequence of RE. Prof A.K.Raychaudhuri \cite{Kar:2006ms} addressed the issue of singularity in early 1950's while he was working on the features of electronic energy bands in metals. The nature of gravitational singularities and generic features of GTR fascinated him. Motivated by cosmology, Raychaudhuri in his seminal paper \cite{Raychaudhuri:1953yv}, \cite{seminal} first pointed that singularity is nothing more than an artifact of the symmetries
	of the matter distribution. He then proposed a time-dependent model of universe without assuming the cosmological principle and its implications on homogeneity and isotropy. The evolution of the
	cosmological expansion ($\Theta$) in a given background is popularly known as RE \cite{Kar:2006ms}.\\
	
	It is to be noted that although RE is a purely geometric identity in Riemannian geometry, yet the effect of gravity comes into picture through the Ricci tensor ($R_{ab}$) projected along the congruence of time-like geodesics. For example when the effect of Einstein gravity (via Einstein's field equations) are fed into the RE together with some physical assumptions on matter content of the Universe, it leads to focusing of geodesic or the FT. Although focusing of congruence leads to formation of congruence singularity yet sometimes with some global assumptions it may lead to cosmological and black-hole singularity. At singularity, all physical laws break down and the space-time geometry becomes pathological. Therefore in order to have a clear picture of the space-time structure these singularities must be tackled more critically. Resolution of singularities has remained puzzle over the decades and attempted in the works \cite{Das:2013oda}-\cite{Chakraborty:2023voy}. It is to be noted that in absence of universally accepted model of accelerated universe, cosmographic quantities namely the expansion scalar ($\Theta$), acceleration etc are emerging as important notions in cosmology and this increases the relevance of RE in cosmology. Thus it is clear that SEC coupled with Einstein's field equations results in Convergence Condition (CC) in GR. Since in other extended theories of gravity the field equations are different so there may arise certain conditions under which the avoidance of singularity is possible. In this context RE and modified CC has been studied in $f(R)$ gravity with inhomogeneous background \cite{Chakraborty:2023ork}, $f(T)$ gravity in homogeneous background \cite{Chakraborty:2023yyz}, scalar tensor theory \cite{Choudhury:2021zij} etc. and certain physical conditions have been determined for the possible avoidance of singularity. This shows that existence of singularity is not a generic one, it depends on the gravity theory and the space-time under consideration. There lies the importance of RE in modified gravity theories in homogeneous and isotropic background. The common link in these works is that all have isotropic background. The anisotropy plays a significant role in early stages of evolution of the universe and hence the study of spatially homogeneous and anisotropic cosmological models is physically and cosmologically significant. Observations state that the Universe is homogeneous and isotropic when the inflationary phase was successfully produced (see \cite{Barrow:1987ia}, \cite{Chakraborty:2023neh} for details of inflation). However CMBR anomalies \cite{Schwarz:2015cma} concluded that there was an anisotropic phase in the early Universe which does not make it exactly uniform. There lies the motivation of the present work where we consider anisotropic Universe described by Kantowski-Sachs (KS) space-time models \cite{Vinutha:2023yee} and explore the effect/role of anisotropy in RE and CC.\\
	
	The exact solutions for homogeneous spacetimes in GR belongs to either Kantowski-Sachs model or the Bianchi Types. KS is the only anisotropic but spatially homogeneous cosmology that does not fall under the Bianchi classification \cite{Leon:2013bra}, \cite{Datta:2021jwr}. These models gained popularity with the publication by Kantowski and Sachs \cite{Kantowski:1966te}. KS space-time has some exciting features. Firstly, its classical and quantum solutions are well known in different contexts \cite{Leon:2010pu}-\cite{Barbosa:2004kp}. Secondly, these models exhibit spherical and transnational symmetry and can be treated as non empty analogs of a part of the extended Schwarzschild metric \cite{Collins:1977fg}. Moreover these models help to study the behavior of the added degrees of freedom in quantum cosmological models. Finally, it may be possible that constructing a KS quantum cosmological model may suggest modifications and adaptions in the quantization methods applied to cosmology. In the present work, anisotropy described by KS model has been analyzed as geometric and physical property of matter. Further which property of anisotropy favors or avoids formation of singularity has also been discussed.
Thus the paper is organized as follows: Section 2 deals with the overview of RE from geometric and physical point of view. Section 3 contains the information regarding Kantowski-Sachs (KS) model and formulation of RE in it. Modified CC has been discussed in Section 4. Section 5 deals with the existence and possible avoidance of singularity in KS model by examining the sign of the Convergence scalar. In section 6, RE in KS model has been made analogous to a differential equation for Harmonic Oscillator and role of anisotropy has been explored in the context of convergence. The paper ends with conclusion and discussion of the obtained results in Section 7.
	\section{Kinematics of flows and Focusing Theorem: Geometry and Physics of RE}
	The RE is associated with the kinematics of flows. Given a vector field, flows are the integral curves generated by that vector field. These curves may be geodesic or non-geodesic in nature. Although the study of geodesics is more relevant and useful when gravity comes into picture. Therefore one may say that a flow can be described by a congruence of time-like or null curves. It is interesting to note that kinematic characteristics of such flows give rise to RE. To be precise, RE is the evolution equation (along the flow) for one of the quantities that characterize the flow in a given space-time background.
	
	In order to define various kinematic quantities that characterize a flow we denote $\tau$ to be the parameter labeling points on the curve in the flow. Further we assume the dimension of the background space-time to be $(n+1)$. Then the gradient of velocity vector field $v^{\mu}$ is a $(0,2)$ tensor and can be expressed as \cite{Kar:2006ms}.
	\begin{equation}
		B_{\mu\nu}=\nabla_{\nu}v_{\mu}=\sigma_{\mu\nu}+\omega_{\mu\nu}+\dfrac{1}{n}\xi_{\mu\nu}\Theta
	\end{equation}
	where
	\begin{equation}
		\sigma_{\mu\nu}=\dfrac{1}{2}\left(\nabla_{\nu}v_{\mu}+\nabla_{\mu}v_{\nu}\right)-\dfrac{1}{n}\xi_{\mu\nu}\Theta\label{eq2}
	\end{equation} is known as shear tensor and $2\sigma^{2}=\sigma_{\mu\nu}\sigma^{\mu\nu}$ is the anisotropy scalar. Further $\sigma_{\mu\nu}$ is the symmetric and traceless part of $B_{\mu\nu}$ i.e,  $\sigma_{\mu\nu}=\sigma_{\nu\mu}$ and $\sigma^{\mu}_{\mu}=0$. It measures anisotropy of the medium. For example in the background of homogeneous and isotropic FLRW space-time, it vanishes. The rotation of the bundle of geodesic or simply the flow is characterized by  \begin{equation}
		\omega_{\mu\nu}=\dfrac{1}{2}\left(\nabla_{\nu}v_{\mu}-\nabla_{\mu}v_{\nu}\right),
	\end{equation}
 the antisymmetric part of $B_{\mu\nu}$ i.e, $\omega_{\mu\nu}=-\omega_{\nu\mu}$ and is known as rotation/ vorticity tensor. Also, $2\omega^{2}=\omega_{\mu\nu}\omega^{\mu\nu}$. Further,
	\begin{equation}
		\xi_{\mu\nu}=g_{\mu\nu}\pm v_{\mu}v_{\nu}
	\end{equation} is the projection tensor/ induced metric (`+' is for time-like curves ($v_{\mu}v^{\mu}=-1$) and `-' is for space-like curves ($v_{\mu}v^{\mu}=+1$)). The expansion scalar or the trace part of $B_{\mu\nu}$ is given by
	\begin{equation}
		\Theta=\nabla_{\mu} v^{\mu}\label{eq5}
	\end{equation} It represents how the geodesic focus/ defocus. The shear, rotation and expansion are associated to the geometry of the cross sectional area enclosing a fixed number of geodesics or a bundle of geodesics orthogonal to the flow lines. The flow in terms of kinematic quantities can be explained as: On moving from one point to another along the flow the shape of this area changes yet it still includes the same bundle of geodesic but may be sheared ($\sigma$ carries the effect), twisted ($\omega$ is responsible) or isotropically larger/smaller (depends on $\Theta$). The evolution equation for $\Theta$ is popularly known as the RE and is given by
	\begin{equation}
		\dfrac{\mathrm{d}\Theta}{\mathrm{d}\tau}=-\dfrac{\Theta^{2}}{3}-2\sigma^{2}+2\omega^{2}-R_{\mu\nu}v^{\mu}v^{\nu}\label{eq6}
	\end{equation} where $R_{\mu\nu}$ is the Ricci tensor projected along the geodesics. This equation governs the rate of $\Theta$ and involves all geometric quantities. Thus RE is simply a geometric identity in Riemannian geometry. However the effect of gravity comes through $R_{\mu\nu}$. In Einstein gravity if the matter satisfies Strong Energy Condition (SEC) then RE leads to Focusing Theorem (FT). This proves the inevitable existence of singularity (not necessarily curvature singularities) in Einstein gravity theory. In the work of Landau and Lifschits \cite{Landau:1975pou} it has been pointed out that a singularity will always imply focusing of geodesics but focusing alone cannot imply a singularity. Thus if one can avoid focusing, formation of singularity might be avoided. In Einstein gravity the last term on the r.h.s of equation (\ref{eq6}) can be written as 
	\begin{equation}
		R_{\mu\nu}v^{\mu}v^{\nu}=(T_{\mu\nu}-\frac{1}{2}Tg_{\mu\nu})v^{\mu}v^{\nu}.
	\end{equation} where $T_{\mu\nu}$ is the energy momentum tensor. Further if the matter content of the universe is normal/usual matter satisfying SEC then
	\begin{equation}
		T_{\mu\nu}v^{\mu}v^{\nu}+\frac{1}{2}T \geqslant0,
	\end{equation}
	Consequently, RE gives \begin{equation}
		\dfrac{\mathrm{d}\Theta}{\mathrm{d}\tau}+\dfrac{1}{3}\Theta^{2} \leqslant0.\label{eq9}
	\end{equation} Integrating the above inequality one gets
	\begin{equation}
		\dfrac{1}{\Theta}\geq \dfrac{1}{\Theta_{0}}+\dfrac{\tau}{3}.\label{eq10*}
	\end{equation} This shows that an initially converging congruence of time-like geodesics begin to focus within a finite value of $\tau$ provided $R_{\mu\nu}v^{\mu}v^{\nu}\geq0$. This is called the FT and $R_{\mu\nu}v^{\mu}v^{\nu}\geq0$ is called Convergence Condition (CC). This is the most vital consequence of the RE and is the key ingredient of the seminal singularity theorems  by Hawking and Penrose. $R_{\mu\nu}v^{\mu}v^{\nu}$ is called the Raychaudhuri scalar and positive semi definiteness of this scalar defines the CC. If under any circumstances we can find that $R_{\mu\nu}v^\mu v^\nu<0$ then we can avoid focusing and formation of singularity. Thus RE not only plays a crucial role in identifying singularity but also suggests its possible removal. There lies the importance of this equation.
	\section{Kantowski-Sachs space-time and RE}
	With an aim to formulate the RE in anisotropic background and to find the effect of anisotropy in CC we consider a general metric for an homogeneous and anisotropic space-time with spatial section topology $\mathbf{R\times S^{2}}$. This is the Kantowski-Sachs (KS) space-time described by the metric \cite{deCesare:2020swb}
	\begin{equation}
		ds^{2}=-dt^{2}+a^{2}(t)dr^{2}+b^{2}(t)(d\theta^{2}+\sin^{2}\theta d\phi^{2})\label{eq11}
	\end{equation} where $a(t)$ and $b(t)$ are two arbitrary and independent functions of cosmic time $t$. The generic form of the energy-momentum tensor in support of this geometry is given by
	\begin{equation}
		T^{\mu}_{\nu}= diag~(-\rho, p_{r},p_{t}, p_{t})\label{eq12}
	\end{equation} where $\rho$ is the energy density of the physical fluid, $p_{r}$ is the radial and $p_{t}$ is the lateral pressure of the physical fluid. The Einstein's field equations defining the above metric (\ref{eq11}) and energy-momentum source (\ref{eq12}) can be written as \cite{Xanthopoulos:1992fh}, \cite{Mendes:1990eb}
	\begin{eqnarray}
		\dfrac{\dot{b}^{2}}{b^{2}}+2\left(\dfrac{\dot{a}}{a}\right)\left(\dfrac{\dot{b}}{b}\right)+\dfrac{1}{b^{2}}=\kappa \rho\label{eq13}\\
		2\dfrac{\ddot{b}}{b}+\left(\dfrac{\dot{b}}{b}\right)^{2}+\dfrac{1}{b^{2}}=-\kappa p_{r}\label{eq14}\\
		\dfrac{\ddot{a}}{a}+\dfrac{\ddot{b}}{b}+\left(\dfrac{\dot{a}}{a}\right)\left(\dfrac{\dot{b}}{b}\right)=-\kappa p_{t}\label{eq15}
	\end{eqnarray} where $\kappa=8\pi G$ is the four dimensional gravitational coupling constant and in units $8 \pi G=1$. Further $p_{r}=\omega_{r}\rho$, $p_{t}=\omega_{t}\rho$ and $\omega_{r}\neq \omega_{t}$ i.e, for the sake of generality we consider distinct EoS for the radial and lateral pressures $p_{r}$ and $p_{t}$ respectively. Thus we introduce metric (\ref{eq12}) and corresponding field equations (\ref{eq13})-(\ref{eq15}) in four dimensions. The average Hubble parameter $H$ in this case is given by
	\begin{equation}
		H=\dfrac{1}{3}\left(\dfrac{\dot{a}}{a}+2\dfrac{\dot{b}}{b}\right)=\dfrac{1}{3}\left(H_{a}+2H_{b}\right)
	\end{equation}
The expansion scalar and anisotropy scalar are given by 
\begin{eqnarray}
	\Theta=\left(\dfrac{\dot{a}}{a}+2\dfrac{\dot{b}}{b}\right)\nonumber\\
		\sigma=\dfrac{1}{\sqrt{3}}\left(\dfrac{\dot{b}}{b}-\dfrac{\dot{a}}{a}\right)=\dfrac{1}{\sqrt{3}}(H_{b}-H_{a})\label{eq17}
	\end{eqnarray} where $H_{a}=\dfrac{\dot{a}}{a}$ and $H_{b}=\dfrac{\dot{b}}{b}$ .  As the present model is Kantowski-Sachs, the metric (\ref{eq11}) is locally rotationally symmetric (LRS) and it belongs to the LRS Class II having rotation tensor $\omega_{ab}=0$ \cite{vanElst:1995eg}, \cite{Marklund:1998sq}. Thus the only non vanishing kinematic quantities are $\Theta$ and $\sigma$. Considering the following transformations:
	\begin{eqnarray}
		V^{3}=ab^{2}\label{eq18}\\
		Z=\dfrac{b}{a}\label{eq19}
	\end{eqnarray}
	 Equations (\ref{eq13}), (\ref{eq18}) and (\ref{eq19}) gives (after simplification)
	\begin{equation}
		3\dfrac{\dot{V}^{2}}{V^{2}}-\dfrac{1}{3}\dfrac{\dot{Z}^{2}}{Z^{2}}+\dfrac{1}{V^{2}Z^{\frac{2}{3}}}=\kappa \rho \label{eq23}
	\end{equation} Using equations (\ref{eq5}) and (\ref{eq18}) one gets
	\begin{equation}
		\Theta=3\dfrac{\dot{V}}{V}\label{eq24}
	\end{equation} and using equations (\ref{eq2}) and (\ref{eq19}) one obtains
	\begin{equation}
		\sigma=\dfrac{1}{\sqrt{3}}\dfrac{\dot{Z}}{Z}\label{eq25}
	\end{equation}
	From equations (\ref{eq23}), (\ref{eq24}) and (\ref{eq25}) one has
	\begin{equation}
		\dfrac{\Theta^{2}}{3}-\sigma^{2}+\dfrac{1}{V^{2}Z^{\frac{2}{3}}}=\kappa \rho\label{eq26}
	\end{equation}
	The measure of acceleration corresponding to the two scale factors $a(t)$ and $b(t)$ in terms of the transformed variables $V$ and $Z$ are
	\begin{eqnarray}
		\dfrac{\ddot{b}}{b}=\dfrac{\ddot{V}}{V}-\dfrac{2}{9}\dfrac{\dot{Z}^{2}}{Z^{2}}+\dfrac{1}{3}\dfrac{\ddot{Z}}{Z}+\dfrac{2}{3}\left(\dfrac{\dot{V}}{V}\right)\left(\dfrac{\dot{Z}}{Z}\right)\label{eq27}\\
		\dfrac{\ddot{a}}{a}=\dfrac{\ddot{V}}{V}+\dfrac{10}{9}\dfrac{\dot{Z}^{2}}{Z^{2}}-\dfrac{2}{3}\dfrac{\ddot{Z}}{Z}-\dfrac{4}{3}\left(\dfrac{\dot{V}}{V}\right)\left(\dfrac{\dot{Z}}{Z}\right)
	\end{eqnarray} Therefore the field equations (\ref{eq14}) and (\ref{eq15}) in terms of the transformed variables can be rewritten as
	\begin{eqnarray}
		2\dfrac{\ddot{V}}{V}+\dfrac{2}{3}\dfrac{\ddot{Z}}{Z}+\dfrac{\dot{V}^{2}}{V^{2}}+2\dfrac{\dot{V}}{V}\dfrac{\dot{Z}}{Z}-\dfrac{1}{3}\dfrac{\dot{Z}^{2}}{Z^{2}}+\dfrac{1}{V^{2}Z^{\frac{2}{3}}}=-\kappa p_{r}\label{eq29}\\
		2\dfrac{\ddot{V}}{V}-\dfrac{1}{3}\dfrac{\ddot{Z}}{Z}+\dfrac{\dot{V}^{2}}{V^{2}}-\dfrac{\dot{V}}{V}\dfrac{\dot{Z}}{Z}+\dfrac{2}{3}\dfrac{\dot{Z}^{2}}{Z^{2}}=-\kappa p_{t}\label{eq30}
	\end{eqnarray}
	Also we have,
	\begin{equation}
		\dfrac{\mathrm{d}\Theta}{\mathrm{d}\tau}+\dfrac{\Theta^{2}}{3}=3\dfrac{\ddot{V}}{V}\end{equation} and,
	\begin{equation}
		\dfrac{\mathrm{d}\sigma}{\mathrm{d}\tau}=\dfrac{1}{\sqrt{3}}\left(\dfrac{\ddot{Z}}{Z}-\dfrac{\dot{Z}^{2}}{Z^{2}}\right)
	\end{equation} 
	By some algebraic manipulation with the field equations we get
	\begin{equation}
		\dfrac{\ddot{Z}}{Z}+3\dfrac{\dot{V}}{V}\dfrac{\dot{Z}}{Z}-\dfrac{\dot{Z}^{2}}{Z^{2}}+\dfrac{1}{V^{2}Z^{\frac{2}{3}}}=\kappa(p_{t}-p_{r})\label{eq33}\end{equation} and,
	\begin{eqnarray}
		\dfrac{\ddot{V}}{V}+\dfrac{2}{9}\dfrac{\dot{Z}^{2}}{Z^{2}}=-\dfrac{\kappa}{6}(\rho+p_{r}+2p_{t})\label{eq34}
	\end{eqnarray}
	Thus the evolution equations for the expansion scalar ($\Theta$) and shear scalar $\sigma$ are given by
	\begin{equation}
		\dfrac{\mathrm{d}\Theta}{\mathrm{d}\tau}+\dfrac{\Theta^{2}}{3}+2\sigma^{2}=-\dfrac{\kappa}{2}(\rho+p_{r}+2p_{t})\label{eq35}
	\end{equation} and,
	\begin{equation}
		\dfrac{\mathrm{d}\sigma}{\mathrm{d}\tau}+\Theta \sigma+\dfrac{\sigma^{2}}{\sqrt{3}}-\dfrac{\Theta^{2}}{3\sqrt{3}}=-\dfrac{\kappa}{\sqrt{3}}(\rho-\frac{1}{3}(p_{t}-p_{r}))\label{eq36}
	\end{equation}
One may note that the above equations (\ref{eq35}) and (\ref{eq36}) are first order, coupled and non-linear in nature. Since the present work is in the background of anisotropic space-time, we have shown the evolution of anisotropy scalar in equation (\ref{eq36}). Further, it is to be noted that the rotation tensor vanishes identically in the present model due to locally rotational symmetry of the KS metric and consideration of hyper-surface orthogonal congruence of time-like geodesic which by virtue of Frobenius theorem implies a zero rotation tensor. However, the evolution of expansion scalar $\Theta$ is of central interest in the context of singularity theorems. Mathematically, the evolution equation for expansion scalar is known as Riccati equation (for ref. see a review of Raychaudhuri equations by Kar and Sengupta \cite{Kar:2006ms}). Historically, the Riccati equation or equation (\ref{eq35}) is known as the Raychaudhuri equation.
	\section{Convergence Condition in KS space-time models}
	In order to study the CC in KS space-time we consider the RE in KS space-time given by equation (\ref{eq35}). Let us write
	\begin{equation}
		\dfrac{\mathrm{d}\Theta}{\mathrm{d}\tau}+\dfrac{\Theta^{2}}{3}=-\tilde{R_{c}}\label{eq37}
	\end{equation}
	where
	\begin{equation}
	\tilde{R_{c}}=\\\dfrac{\kappa}{2}(\rho+p_{r}+2p_{t})+2\sigma^{2}\label{eq38}
	\end{equation}
 Here we consider $p_{r}=\omega_{r}\rho$ and $p_{t}=\omega_{t}\rho$, $\omega_{r}$ and $\omega_{t}$ are the radial and transverse EoS parameters. From the modified form of the RE (\ref{eq37}) the CC now becomes $\tilde{R_{c}}\geq0$. Thus the possibilities of focusing are listed as follows:
	\begin{itemize}
		\item $\dfrac{\kappa}{2}(\rho+p_{r}+2p_{t})\geq0$ i.e, matter satisfies SEC. In this case $	\tilde{R_{c}}\geq0$ (follows from equation \ref{eq38}). Since the positive semi definiteness of $\tilde{R_{c}}$ leads to convergence of the bundle of geodesics therefore we name it as Convergence Scalar and we shall use this term in the following sections.
		\item Matter violates SEC i.e, $\dfrac{\kappa}{2}(\rho+p_{r}+2p_{t})<0$ but $2\sigma^{2}>\dfrac{\kappa}{2}|(\rho+p_{r}+2p_{t})|$. 
	\end{itemize}
	Based on the above discussion we conclude that the anisotropy term favors the convergence/ formation of congruence singularity. However this term alone does not lead to CC. If the matter is attractive in nature i.e, $1+\omega_{r}+2\omega_{t}\geq0$ then CC is automatically satisfied. However if the matter is repulsive in nature i.e, $1+\omega_{r}+2\omega_{t}<0$ then anisotropy can not alone lead to CC but needs an extra condition namely $2\sigma^{2}>\dfrac{\kappa}{2}|(\rho+p_{r}+2p_{t})|$. Thus in comparison to FLRW model the anisotropy term may be interpreted as a matter part which is attractive in nature. This interpretation can be justified from Einstein's idea of gravity where geometry and matter are equivalent quantities. Thus to avoid formation of singularity in KS space-time we need $\dfrac{\kappa}{2}(\rho+p_{r}+2p_{t})<0$ and $2\sigma^{2}<\dfrac{\kappa}{2}|(\rho+p_{r}+2p_{t})|$. In other words, to avoid focusing matter cannot be usual in nature. Further violation of SEC must be dominant over the anisotropy term in order to avoid singularity formation. Hence with usual matter singularity is inevitable in a general KS model. \\  \\
	Thus the general KS space-time classically has a past and a future singularity, which can be an anisotropic structure such as a barrel, cigar, a pancake or an isotropic point like structure depending on the initial conditions on anisotropic shear and matter \cite{Ghorani:2021xrs}. Evolution of geodesics terminates at these classical singularities which is identified by the divergence of expansion and shear scalars in the presence of matter (which contributes to the energy density). Existence of singularities pushes GR to the limits of its validity and hence a quantum gravitational treatment which becomes dominant in strong gravity regimes may alleviate the classical singularity.
	\section{Singularity analysis in a constrained KS model}
	In order to examine the existence and possible avoidance of singularity we shall study the sign of the Convergence Scalar $R_{c}$ in this section. We know that $R_{c}\geq0$ ensures convergence i.e, it is the condition for focusing. Also it is known that if a space time has a singularity, it means that a bundle of geodesic will tend to focus at the singularity. Thus if we can avoid the focusing of geodesic by making $R_{c}<0$, we can avoid the formation of a singularity in a classical space-time. This is because if there were a singularity, focusing would have inevitably happened there. In this section we find the physical conditions under which $R_{c}$ can be made negative.  For this we consider the anisotropic relation i.e, the physical condition that the expansion scalar $\Theta(t)$ is proportional to the shear scalar $\sigma(t)$ i.e, $\Theta\propto\sigma$ \cite{Kantowski:1966te}, \cite{Sa:1992aw}- \cite{Koussour:2022jss}. This results in the relation between the scale factors as
	\begin{equation} a=b^{m} \label{eq39}
		\end{equation}where $m$ is an arbitrary real number and $m\neq0,1$ to ensure non triviality and anisotropy. Thorne \cite{Thorne} justified this physical law based on the observations of the velocity redshift relation for extragalactic sources which suggest that the Hubble expansion of the Universe is isotropic at present time with 30$\%$ \cite{Kantowski:1966te}. More precisely, the redshift studies put the limit $\dfrac{\sigma}{\Theta}\leq0.3$, the ratio of the shear to expansion scalar in the vicinity of our galaxy at present time. Further it was pointed out by Collins et al. \cite{Collins:1977fg} that the normal congruence to the homogeneous expansion for spatially homogeneous metric satisfies the condition $\dfrac{\sigma}{\Theta}=$ constant. Bunn et al. \cite{Bunn:1996ut} did statistical analysis on 4-yr CMB data and set a limit for primordial anisotropy to be less than $10^{-3}$ in Planck epoch. In Literature many researchers have thus used this condition motivated by the above physical implications while dealing with KS space-time models. This is the physical motivation behind choosing this constraint. Now using equations (\ref{eq17}) and (\ref{eq39}) we have
	\begin{equation}
		2\sigma^{2}=\dfrac{2}{3}(1-m)^{2}\left(\dfrac{\dot{b}}{b}\right)^{2}\label{eq40}
	\end{equation}
From equation (\ref{eq40}) and (\ref{eq38}) the expression for $\tilde{R_{c}}$ is given by
\begin{equation}
		\tilde{R_{c}}=\\\dfrac{\kappa}{2}(\rho+p_{r}+2p_{t})+\dfrac{2}{3}(1-m)^{2}\left(\dfrac{\dot{b}}{b}\right)^{2}\\
	=\dfrac{\kappa\rho}{2}(1+\omega_{r}+2\omega_{t})+\dfrac{2}{3}(1-m)^{2}\left(\dfrac{\dot{b}}{b}\right)^{2}\label{eq41}
	\end{equation}
Let us assume,
\begin{equation}
	\omega_{r}=\alpha~\omega_{t}\label{eq42}
\end{equation} where $\alpha\neq1$ to maintain anisotropy.
Finally by doing some algebraic manipulation with the field equations (\ref{eq13})-(\ref{eq15}) (to eliminate $\dot{b}$ and $b$) and using equations (\ref{eq39}), (\ref{eq41}), expression for $\tilde{R_{c}}$ takes the form
\begin{equation}
		\tilde{R_{c}}=\kappa\rho R_{c}
	\end{equation}
where
\begin{equation}
	R_{c}=\left[\left(\alpha+\dfrac{1}{2}\right)+\dfrac{(1-m)^{2}}{3m}\left(1-\dfrac{2\alpha}{m+1}\right)\right]\omega_{r}+\dfrac{2(1-m)^{2}}{3m(m+1)}+\dfrac{1}{2}+\dfrac{(1-m)^{3}}{3m(m+1)}
	\end{equation}\\ \\
Thus the sign of $\tilde{R_{c}}$ depends upon the sign of $R_{c}$ provided $\kappa\rho>0$. We assume that $\kappa\rho>0$ and plot $R_{c}$ w.r.t $\omega_{r}$ and $m$ (the power appears in equation (\ref{eq39})) in the following FIG.\ref{f1}. We consider $\omega_{r}\in(-1,1)$ and $m\in(-10,-5)$ or $m\in(2,10)$ and see the variation of $R_{c}$ for different values of $\alpha$. Based on the graphs the range/ values of $\alpha$, $m$ and $\omega_{r}$ are represented in a tabular form in tables (\ref{t1}) and (\ref{t2}) for which $R_{c}<0$.\\ \\ 
	We observed that focusing theorem  (\ref{eq10*}) follows from the evolution equation for expansion (\ref{eq35}) by assuming SEC on matter. Since in the previous section we found that shear is in favor of focusing, this motivates us to deduce the mathematical statement of focusing theorem or whether it actually holds in terms of $\sigma$. However the general evolution equation for shear (\ref{eq36}) is highly non-linear and coupled posing difficulties to fulfill the aim. Therefore we attempt to see whether the focusing theorem holds following the evolution of shear in the physically motivated constrained KS model (\ref{eq39}). For this we need the explicit relation between $\Theta$ and $\sigma$.
It is to be noted that the relation between $\Theta$ and $\sigma$ for the choice in equation (\ref{eq39}) can be explicitly written as
\begin{equation}
	\Theta=\pm 3\sqrt{3}\left(\dfrac{m+2}{m-1}\right)\sigma\label{eq45*}
\end{equation} 
 The above relation follows from equations (\ref{eq18}), (\ref{eq19}), (\ref{eq24}), (\ref{eq25}) and (\ref{eq39}).
Hence, the evolution equation for shear follows from eq.(\ref{eq35}) and eq.(\ref{eq45*}) as
\begin{equation}
	\dfrac{d\sigma}{d\tau}+\beta\sigma^{2}=-\left(\dfrac{(m-1)}{3\sqrt{3}(m+2)}\right)\dfrac{\kappa}{2}\left(\rho+p_{r}+2p_{t}\right)\label{eq46*}
\end{equation} where
\begin{equation}
	\beta=\dfrac{\left(\left(\frac{m+2}{m-1}\right)^{2}+2\right)(m-1)}{3\sqrt{3}(m+2)}
\end{equation} if $	\Theta=+3 \sqrt{3}\left(\dfrac{m+2}{m-1}\right)\sigma$. 
Now considering expanding model of universe i.e, $m>1$ and matter, usual in nature which satisfies the SEC ( $\dfrac{\kappa}{2}\left(\rho+p_{r}+2p_{t}\right)\geq0$ ) then eq.(\ref{eq46*}) gives rise to the inequality
\begin{equation}
	\dfrac{1}{\sigma}\geq \beta\tau+\dfrac{1}{\sigma_{0}}
\end{equation} where
$\sigma_{0}$ is the constant of integration and physically $\sigma_{0}=\sigma(\tau=0)$.  We do not interpret this result physically since the choice $\Theta=+3\sqrt{3}\left(\dfrac{m+2}{m-1}\right)\sigma$ is not physically feasible for $m>1$. This is because $\Theta\rightarrow-\infty$ implies convergence which will make $\sigma\rightarrow-\infty$. However anisotropy is in favor of convergence which necessitates $\sigma\rightarrow+\infty$ for convergence. On the other hand if we consider the physically feasible choice $\Theta=-3\sqrt{3}\left(\dfrac{m+2}{m-1}\right)\sigma$, then we have
\begin{equation}
	\dfrac{d\sigma}{d\tau}-\beta\sigma^{2}=\left(\dfrac{(m-1)}{3\sqrt{3}(m+2)}\right)\dfrac{\kappa}{2}\left(\rho+p_{r}+2p_{t}\right)
\end{equation}
 Assuming an expanding universe with matter satisfying SEC we arrive at the inequality
 \begin{equation}
 	\dfrac{1}{\sigma}\leq-\beta\tau+\dfrac{1}{\sigma_{0}}\label{eq50**}
 \end{equation}
Recall the definition of $\Theta$ which is  the rate of change of the cross-sectional area orthogonal to the bundle of geodesics. Thus $\Theta\rightarrow+\infty$ implies a divergence of the bundle while $\Theta\rightarrow-\infty$ implies a complete convergence. Also, for convergence there must be a negative expansion initially. Therefore in this particular case with $\Theta=-3\sqrt{3}\left(\dfrac{m+2}{m-1}\right)\sigma$, as $\Theta_{0}<0$ we have $\sigma_{0}>0$ for expanding model with $m>1$. Inequality (\ref{eq50**}) shows that an initially converging bundle of geodesic will diverge after some finite time $\tau$. Therefore in this constrained KS model with $\Theta=-3\sqrt{3}\left(\dfrac{m+2}{m-1}\right)\sigma$, although there is an initial singularity but in the course of evolution of the universe there may not be any future singularity in the sense that focusing does not happen and it is reflected in the inequality (\ref{eq50**}). However the possible resolution of the initial singularity is attempted via the signature of $R_{c}$ in FIG. (\ref{f1}).
\begin{table}[h!]
\begin{center}
	\begin{tabular}{ | c| c| c| } 
		\hline
	\multirow{1}{*}{$\alpha=1.1$}& $\omega_{r}\in[-1,-0.5]$ & $m\in[2,10]$ \\ 
		\hline
			\multirow{1}{*}{$\alpha=1.1$}& $\omega_{r}$ close to $+1$ & higher negative power of $m$ $(m\rightarrow-10)$ \\ 
			\hline
		$\alpha=0.1$ & $\omega_{r}$ close to $-1$ & $m\in[2,10]$ \\ 
		\hline
			$\alpha=0.1$ & $\omega_{r}$ close to $+1$ & $m\in[-10,-5]$ \\ 
		\hline
		$\alpha=1.5$ & $\omega_{r}\in[-1,-0.5]$ & $m\in[2,10]$ \\ 
		\hline
			$\alpha=1.5$ & $\omega_{r}$ close to $+1$ & $m\in[-10,-5]$ \\ 
		\hline
		$\alpha=2.0$ & $\omega_{r}$ close to $-1$ & $m\in[2,10]$ \\
			\hline
				$\alpha=2.0$ & $\omega_{r}$ close to $+1$ & $m\in[-10,-5]$ \\ 
			\hline
				$\alpha=0.9$ & $\omega_{r}\in[-1,-0.5]$ & $m\in[2,10]$ \\
				\hline
					$\alpha=0.9$ & $\omega_{r}$ close to $+1$ & $m\in[-10,-5]$ \\ 
				\hline
	\end{tabular}
\end{center}
	\caption{Table showing the positive values of $\alpha$ and corresponding range of $\omega_{r}$ and $m$ which make $R_{c}<0$}\label{t1}
\end{table}
\begin{table}[h!]
	\begin{center}
		\begin{tabular}{ | c | c| c | } 
			\hline
			\multirow{1}{*}{$\alpha=-1$}& $\omega_{r}$ close to $-1$ &  higher positive power of $m$ $(m\rightarrow 10)$ \\ 
			\hline
				\multirow{1}{*}{$\alpha=-1$}& $\omega_{r}$ close to $+1$ &  higher negative power of $m$ $(m\rightarrow -10)$ \\ 
			\hline
			$\alpha=-1.5$ & $\omega_{r}\in [-1,-0.5] $&  higher positive power of $m$ $(m\rightarrow 10)$\\ 
			\hline
				$\alpha=-1.5$ & $\omega_{r}\in [0.5,1] $&  higher negative power of $m$ $(m\rightarrow -10)$\\ 
			\hline
			$\alpha=-2$ & $\omega_{r}\in[-1,-0.5]$ & higher positive power of $m$ $(m\rightarrow 10)$\\ 
			\hline
			$\alpha=-2$ & $\omega_{r}\in[0.5,1]$ & higher negative power of $m$ $(m\rightarrow -10)$\\ 
			\hline
		\end{tabular}
	\end{center}
	\caption{Table showing the negative values of $\alpha$ and corresponding range of $\omega_{r}$ and $m$ which make $R_{c}<0$}\label{t2}
\end{table}
\newpage
 The two tables Table.\ref{t1} and Table.\ref{t2} cover up all possibilities which make $R_{c}<0$ as depicted in graphs. From Table.\ref{t1}, we find that when $\alpha=0.9/1.1/1.5$ (closer to $+1$) i.e, a little effect of anisotropy and positive value of $m$ (which accounts for expanding model since $a=b^{m}$, $m>0$) is taken into consideration then both cosmological constant having EoS parameter = $-1$ and fluid having EoS parameter $-1\leq\omega_{r}<-\dfrac{1}{3}$ may avoid singularity. Such type of fluids are non phantom in nature. Also according to the observations, the expansion of the universe is accelerating (both $\Theta$ and $\dot{\Theta}>0$) for any EoS$<-\dfrac{1}{3}\in[-1,-0.5]$ and this does not allow geodesics to focus by ensuring $R_{c}<0$. Thus our results are at par with the observations with little effect of anisotropy which may be considered as comparable to isotropic universe. On the other hand, if we substantially increase the effect of anisotropy by choosing $\alpha=0.1/2.0$ (far from $+1$) then avoidance of singularity is guaranteed for an expanding model ($m>0$) if and only if the EoS approaches to that of Cosmological Constant. However, if we consider $m<0$ i.e, as $b(t)$ increases $a(t)$ will decrease then whatever be the effect of anisotropy a stiff fluid (having EoS parameter close to $+1$) is always able to avoid the singularity formation.\\
 From Table.\ref{t2}, we find that both negative values of $\alpha$ and $m$ (higher negative powers) may avoid singularity formation provided the fluid is a stiff fluid or a fluid that has a positive EoS $\in[0.5,1]$. On the other hand whatever be the value of $\alpha<0$, either cosmological constant or fluid having EoS $\in[-1,-\dfrac{1}{3}]$ may avoid the singularity formation in an expanding universe (provided $m$ is highly positive).
\section{Role of Anisotropy in convergence: A Harmonic Oscillator approach}
In this section we aim to show how the RE which is a first order non linear differential equation can be converted to a second order Hill-Type equation or differential equation for a Harmonic Oscillator. For this, we eliminate $2\sigma^{2}$ between equations (\ref{eq26}) and (\ref{eq35}) so that one has
\begin{equation}
	\dfrac{d\Theta}{d\tau}+\Theta^{2}+\dfrac{2}{V^{2}Z^{\frac{2}{3}}}=\dfrac{\kappa\rho}{2}\left(5+\omega_{r}+2\omega_{t}\right)\label{eq45}
\end{equation}
Now we consider a transformation 
\begin{equation}
	\Theta=\dfrac{d\ln Y}{d\tau}=\dfrac{1}{Y}\dfrac{dY}{d\tau}
\end{equation}
Under this transformation equation (\ref{eq45}) can be written as
\begin{equation}
	\dfrac{d^{2}Y}{d\tau^{2}}+\left(\dfrac{2}{V^{2}Z^{\frac{2}{3}}}-\dfrac{\kappa\rho}{2}(5+\omega_{r}+2\omega_{t})\right)Y=0\label{eq47}
\end{equation}
Thus, equation (\ref{eq47}) can be identified as a Hill-Type equation or the differential equation for a harmonic oscillator in the transformed variable $Y$ with time varying frequency $\omega_{0}$ where
\begin{equation}
	\omega_{0}^{2}=\left(\dfrac{2}{V^{2}Z^{\frac{2}{3}}}-\dfrac{\kappa\rho}{2}(5+\omega_{r}+2\omega_{t})\right)\geq0\label{eq48}
\end{equation}
Now we show how equation (\ref{eq47}) can be used to deduce the criterion for convergence. To do so, we recall the physical meaning of $\Theta$ (by Kar and SenGupta \cite{Kar:2006ms}). $\Theta$ is nothing but the rate of change of the cross-sectional area orthogonal to the bundle of geodesics. Thus $\Theta\rightarrow-\infty$ implies a convergence of the bundle while $\Theta\rightarrow+\infty$ implies a complete divergence. For convergence, there must be a negative expansion initially. Finally with $\dot{Y}<0$ we should end up at a root of $Y$ at some finite time say, $\tau$ to have a negatively infinite expansion. Thus the requirement for convergence reduces to the criterion for the existence of roots of $Y$ at some finite value of $\tau$. This can be linked with the famous Strum-Comparison theorem of second order differential equations which necessitates $\left(\dfrac{2}{V^{2}Z^{\frac{2}{3}}}-\dfrac{\kappa\rho}{2}(5+\omega_{r}+2\omega_{t})\right)\geq0$ to be the condition for convergence or simply the Convergence Condition (CC).\\ \\
Using equations (\ref{eq24}) and (\ref{eq25}) in equation (\ref{eq45*}) we have
\begin{equation}
	\dfrac{\dot{V}}{V}=\pm \left(\dfrac{m+2}{3(m-1)}\right)\dfrac{\dot{Z}}{Z}
\end{equation}
Solving it we get
\begin{equation}
	V=V_{0}Z^{\pm\frac{(m+2)}{3(m-1)}}
\end{equation} with $V_{0}$, a constant of integration.
Thus, the first term on the r.h.s of equation (\ref{eq48}) can be written as
\begin{equation}
\dfrac{2}{V^{2}Z^{\frac{2}{3}}}=\dfrac{2}{V_{0}Z^{\frac{2}{3}\left(\frac{m+2}{m-1}\right)}}\label{eq53}
\end{equation} if positive sign is considered on the r.h.s of equation  (\ref{eq45*}) and,
\begin{equation}
	\dfrac{2}{V^{2}Z^{\frac{2}{3}}}=\dfrac{2}{V_{0}Z^{\frac{2}{1-m}}}\label{eq54}
\end{equation} if the corresponding sign is negative.
It is to be noted from equations (\ref{eq25}), (\ref{eq53}) and (\ref{eq54}) that the first term on the r.h.s of equation (\ref{eq48}) is indicative of shear/ anisotropy. On the other hand if we look at the second term on the r.h.s of equation (\ref{eq48}) anisotropy occurs for $\omega_{r}\neq \omega_{t}$ which is true according to our assumption in equation (\ref{eq42}) . Further if we look at the CC, we find that anisotropy represented by the term $\dfrac{2}{V^{2}Z^{\frac{2}{3}}}$ is in favor of convergence while anisotropy represented by the term $\dfrac{\kappa\rho}{2}(5+\omega_{r}+2\omega_{t})$ is against convergence if the matter satisfies SEC ($\rho(1+\omega_{r}+2\omega_{t})\geq0$) i.e, in case of usual matter. Based on the above discussion, we can interpret these two types of anisotropy as geometric and physical anisotropy so that the former assists convergence while the later opposes it provided, matter is attractive in nature. Thus the above analysis shows the role of anisotropy in convergence explicitly.
\begin{figure}[h!]
	\begin{minipage}{0.3\textwidth}
		\centering\includegraphics[height=6cm,width=6.5cm]{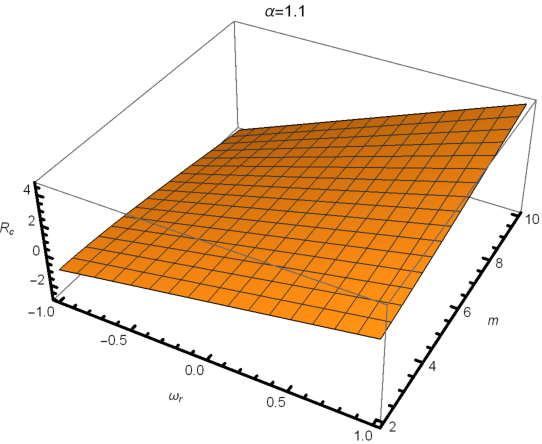}
	\end{minipage}~~~~
	\begin{minipage}{0.3\textwidth}
		\centering\includegraphics[height=6cm,width=6.5cm]{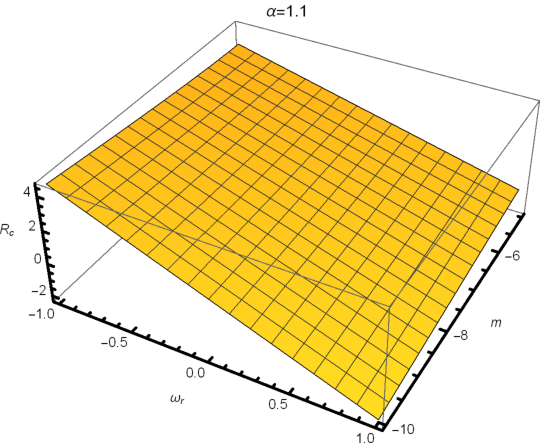}
	\end{minipage}\hfill
	\begin{minipage}{0.3\textwidth}
		\centering\includegraphics[height=6cm,width=7cm]{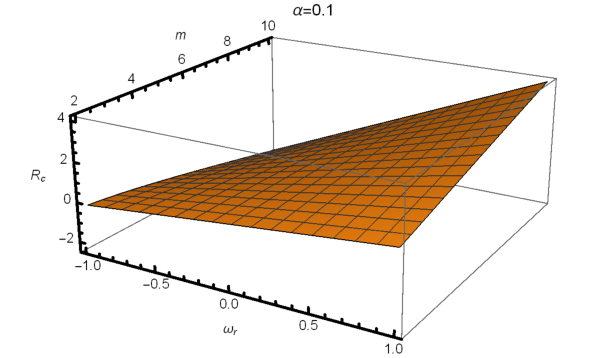}
	\end{minipage}~~~~~~~~~~~~~~~~~~~~~~~~~~
\begin{minipage}{0.3\textwidth}
	\centering\includegraphics[height=6cm,width=6.5cm]{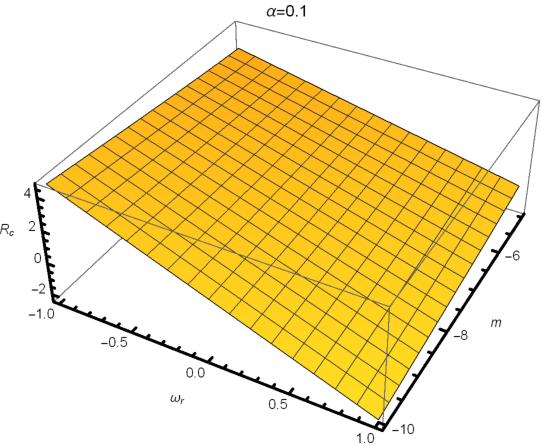}
\end{minipage}
	\begin{minipage}{0.3\textwidth}
	\centering\includegraphics[height=6cm,width=6.5cm]{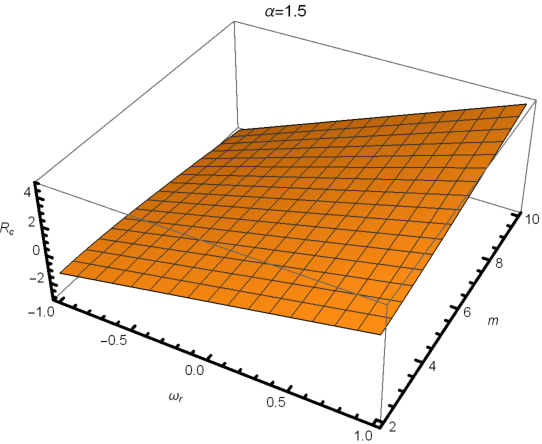}
\end{minipage}\hfill
\begin{minipage}{0.3\textwidth}
	\centering\includegraphics[height=6cm,width=6.5cm]{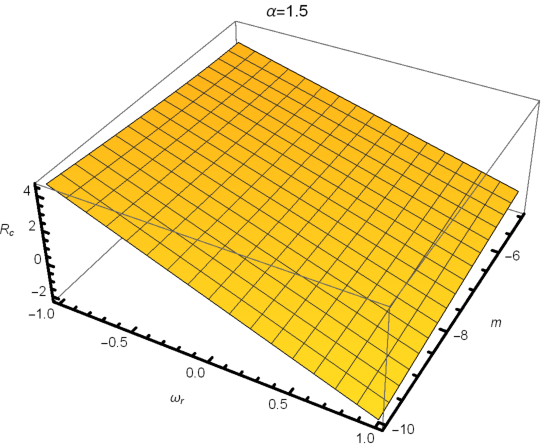}
\end{minipage}
\begin{minipage}{0.3\textwidth}
	\centering\includegraphics[height=6cm,width=6.5cm]{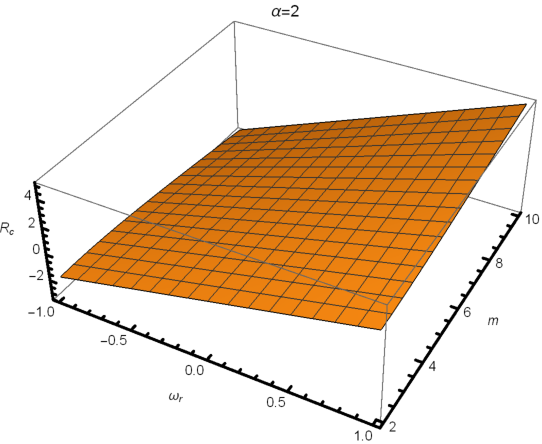}
\end{minipage}~~~~~~~~~~~~~~~~~~~~~~~~~~~~~~~~~~~~~~~~`
\begin{minipage}{0.3\textwidth}
	\centering\includegraphics[height=6cm,width=8cm]{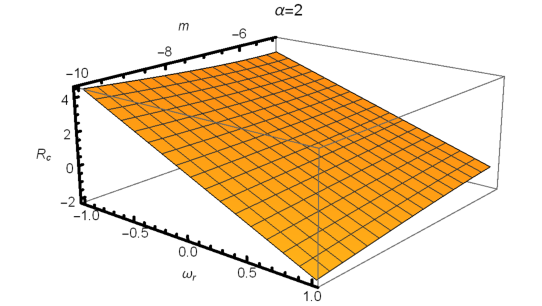}
\end{minipage}
\end{figure}
\begin{figure}
\begin{minipage}{0.3\textwidth}
	\centering\includegraphics[height=6cm,width=7cm]{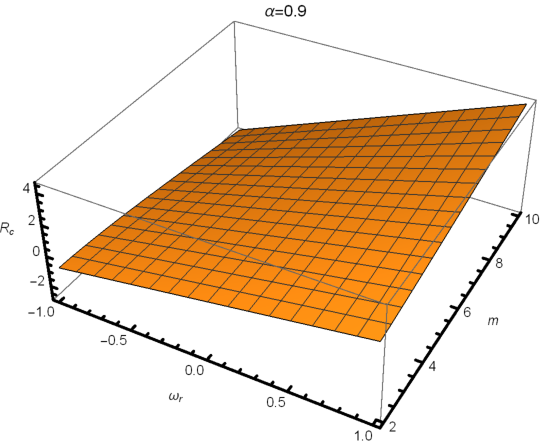}
\end{minipage}~~~~
\begin{minipage}{0.3\textwidth}
	\centering\includegraphics[height=5.8cm,width=6.8cm]{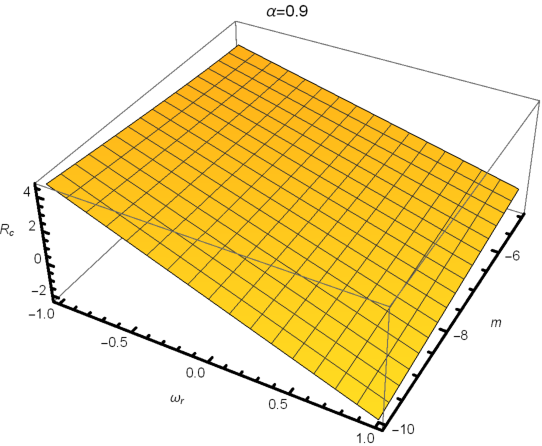}
\end{minipage}
\begin{minipage}{0.3\textwidth}
	\centering\includegraphics[height=6cm,width=7cm]{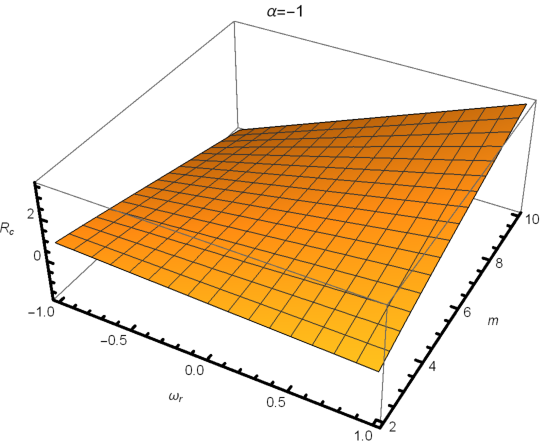}
\end{minipage}~~~~~~~~~~~~~~~~~~~~~~~~~~~~~~~~~~~~~~~~~~~
\begin{minipage}{0.3\textwidth}
	\centering\includegraphics[height=6cm,width=7cm]{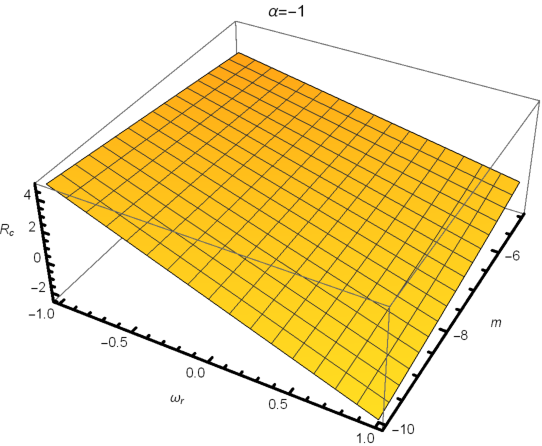}
\end{minipage}
\begin{minipage}{0.3\textwidth}
	\centering\includegraphics[height=6cm,width=7cm]{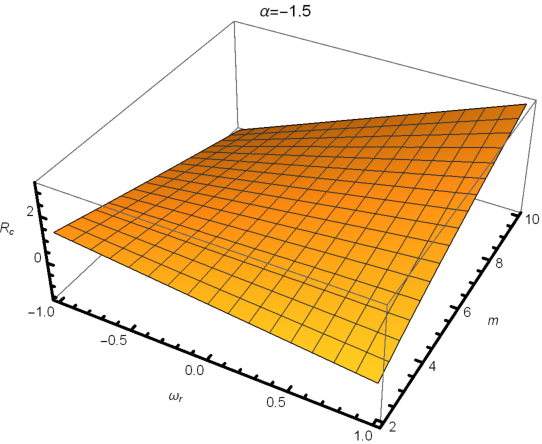}
\end{minipage}~~~~~~~~~~~~~~~~~~~~~~~~~~~~~~~~~~~~~~~~~~~~~~~
\begin{minipage}{0.3\textwidth}
	\centering\includegraphics[height=6cm,width=7cm]{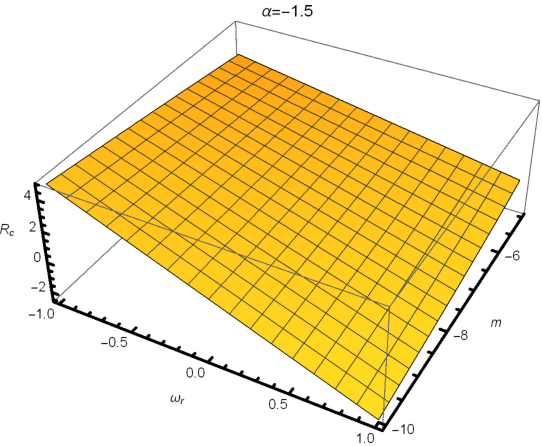}
\end{minipage}
\begin{minipage}{0.3\textwidth}
	\centering\includegraphics[height=6cm,width=7cm]{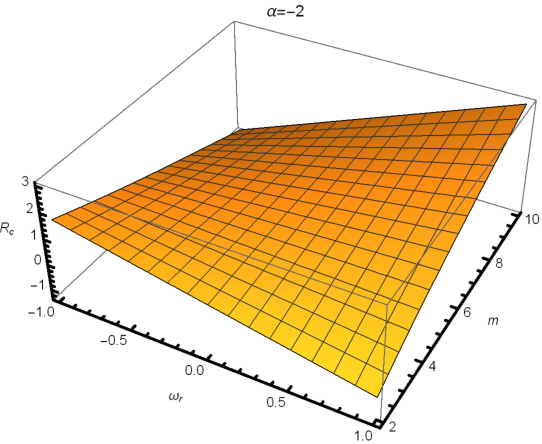}
\end{minipage}~~~~~~~~~~~~~~~~~~~~~~~~~~~~~~~~~~~~~~~~~~~~~~~
\begin{minipage}{0.3\textwidth}
	\centering\includegraphics[height=6cm,width=7cm]{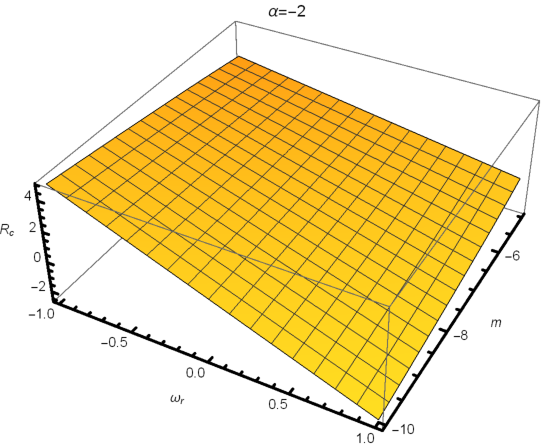}
\end{minipage}
\caption{\small Graphs showing the variation of $R_{c}$ with $m$ and $\omega_{r}$ for arbitrary $\alpha$ (fixed after choice)}\label{f1}
\end{figure}
	\section{Conclusion and Discussion}
	The present work vividly describes the formulation of Raychaudhuri equation (evolution of expansion) in the background of homogeneous and anisotropic Kantowski Sachs space-time model. An extensive analysis of the kinematic quantities that appear in the RE has been carried out from purely geometric point of view. The evolution of shear has been found to be a first order non linear coupled differential equation in a general KS metric. Although Einstein's GR is the most successful theory of gravity to describe physical reality, yet the Focusing theorem which follows as a consequence of RE is the turning point in GR. This is because it hints the inevitable existence of singularity in a classical space-time. Focusing or simply the convergence of a bundle of geodesic requires a Convergence condition (CC) to happen. Thus in the present work we attempted to study the role of anisotropy in focusing both in a general KS model as well as in a physically motivated constrained KS model. \\ \\
	It has been found that a general KS space-time has an initial singularity and anisotropy assists the convergence in the presence of usual matter. However with repulsive matter anisotropy alone can not lead to convergence but needs some extra condition related to the matter part. Therefore as compared to FLRW class of models, anisotropy may be considered as a matter part which is attractive in nature and hence facilitates convergence. It has also been shown following the evolution equation for shear that focusing does not occur in some finite time in the future. However, the constrained KS model may avoid the initial singularity with suitable choice of the parameters involved. For example, an expanding universe with either cosmological constant or non-phantom energy may avoid the formation of singularity in the presence of anisotropy. Although there are possibilities to avoid singularity with negative values of $\alpha$ and $m$ mathematically, yet we avoid discussing them for their physical irrelevance.\\ \\
	The paper also gives a transformation under which the RE which is a first order non linear differential equation in $\Theta$ can be converted to a second order differential equation in a transformed variable $Y$. The second order differential equation so formed is analogous to the evolution equation for a harmonic oscillator. CC has been restated in this context from the physical definition of $\Theta$ along with some initial assumptions and it is found that the CC is associated with the time varying frequency of the oscillator. Explicit expression for the frequency of the oscillator shows the dependence of anisotropy. This analysis further points out a two fold feature of anisotropy namely geometric and physical anisotropy  of which the former assists convergence while the latter defies it. Thus the paper studies the consequences of RE and corresponding convergence condition via the signature of the convergence scalar in the presence of anisotropy and brings out an interesting dual behavior of anisotropy towards convergence via a Harmonic oscillator approach.
	\section*{Acknowledgment}
	The authors thank the anonymous reviewers for their valuable suggestion which led to improvement of the manuscript. M.C thanks University Grant Commission (UGC) for providing the Junior Research Fellowship(ID:211610035684/JOINT CSIR-UGC NET JUNE-2021). S.C. thanks FIST program of DST,
	 Department of Mathematics, JU(SR/FST/MS-II/2021/101(C)).
	

\begin{thebibliography}{100}
			\bibitem{Wald:1984rg}
		R.~M.~Wald,
		``General Relativity,''
		Chicago Univ. Pr., 1984
		\bibitem{Weinberg:1972kfs}
		S.~Weinberg,
		``Gravitation and Cosmology: Principles and Applications of the General Theory of Relativity,''
		John Wiley and Sons, 1972
		\bibitem{LIGOScientific:2017vwq}
		B.~P.~Abbott \textit{et al.} [LIGO Scientific and Virgo],
		Phys. Rev. Lett. \textbf{119}, no.16, 161101 (2017)
		\bibitem{LIGOScientific:2016aoc}
		B.~P.~Abbott \textit{et al.} [LIGO Scientific and Virgo],
		Phys. Rev. Lett. \textbf{116}, no.6, 061102 (2016)
			\bibitem{Hawking:1973uf}
		S.~W.~Hawking and G.~F.~R.~Ellis,
		Cambridge University Press, 2011,
		\bibitem{Penrose:1964wq}
		R.~Penrose,
		Phys. Rev. Lett. \textbf{14}, 57-59 (1965)
		
		\bibitem{Hawking:1970zqf}
		S.~W.~Hawking and R.~Penrose,
		Proc. Roy. Soc. Lond. A \textbf{314}, 529-548 (1970)
			\bibitem{Raychaudhuri:1953yv}
		A.~Raychaudhuri,
		Phys. Rev. \textbf{98}, 1123-1126 (1955)
		\bibitem{Burger:2018hpz}
		D.~J.~Burger, N.~Moynihan, S.~Das, S.~Shajidul Haque and B.~Underwood,
		Phys. Rev. D \textbf{98}, no.2, 024006 (2018)
		
		\bibitem{Kar:2006ms}
		S.~Kar and S.~SenGupta,
		Pramana \textbf{69}, 49 (2007)
		\bibitem{Landau:1975pou}
		L.~D.~Landau and E.~M.~Lifschits,
		Pergamon Press, 1975,
		ISBN 978-0-08-018176-9
		\bibitem{Ehlers:2006aa}
		J.~Ehlers,
		Int. J. Mod. Phys. D \textbf{15}, 1573-1580 (2006).
		\bibitem{Kar:2008zz}
		S.~Kar,
		Resonance J. Sci. Educ. \textbf{13}, 319-333 (2008).
		\bibitem{Horwitz:2021lyc}
		L.~P.~Horwitz, V.~S.~Namboothiri, G.~Varma K, A.~Yahalom, Y.~Strauss and J.~Levitan,
		Symmetry \textbf{13}, no.6, 957 (2021)
		
		\bibitem{Dadhich:2005qr}
		N.~Dadhich,
		``Derivation of the Raychaudhuri equation,''18 Nov (2022)
		https://arxiv.org/abs/gr-qc/0511123
		\bibitem{Borde:1987qr}
		A.~Borde,
		Class. Quant. Grav. \textbf{4}, 343-356 (1987)
		doi:10.1088/0264-9381/4/2/015
		\bibitem{Albareti:2012se}
		F.~D.~Albareti, J.~A.~R.~Cembranos and A.~de la Cruz-Dombriz,
		JCAP \textbf{12}, 020 (2012)
		\bibitem{Chakraborty:2023lav}
		M.~Chakraborty and S.~Chakraborty,
		[arXiv:2308.10498 [gr-qc]].
		\bibitem{seminal}
		Krori, K.D. \& Dutta, Sabita \& Das, Kanika \& Mahanta, Chandra, Indian Journal Of Physics 
		82(5)531-537 (2008). 
		\bibitem{Das:2013oda}
		S.~Das,
		Phys. Rev. D \textbf{89}, no.8, 084068 (2014)
		\bibitem{Ali:2014qla}
		A.~F.~Ali and S.~Das,
		Phys. Lett. B \textbf{741}, 276-279 (2015)
		\bibitem{Blanchette:2021vid}
		K.~Blanchette, S.~Das and S.~Rastgoo,
		JHEP \textbf{09}, 062 (2021)
		\bibitem{Burger:2018hpz}
		D.~J.~Burger, N.~Moynihan, S.~Das, S.~Shajidul Haque and B.~Underwood,
		Phys. Rev. D \textbf{98}, no.2, 024006 (2018)
		\bibitem{Blanchette:2020kkk}
		K.~Blanchette, S.~Das, S.~Hergott and S.~Rastgoo,
		Phys. Rev. D \textbf{103}, no.8, 084038 (2021)
		\bibitem{Chakraborty:2023voy}
		M.~Chakraborty and S.~Chakraborty,
		Annals Phys. \textbf{457}, 169403 (2023)
		\bibitem{Chakraborty:2023ork}
		M.~Chakraborty, A.~Bose and S.~Chakraborty,
		Phys. Scripta \textbf{98}, no.2, 025007 (2023)
		\bibitem{Chakraborty:2023yyz}
		M.~Chakraborty and S.~Chakraborty,
		Class. Quant. Grav. \textbf{40}, no.15, 155010 (2023)
		\bibitem{Choudhury:2021zij}
		S.~G.~Choudhury, A.~Dasgupta and N.~Banerjee,
		Int. J. Geom. Meth. Mod. Phys. \textbf{18}, no.08, 2150115 (2021)
		\bibitem{Barrow:1987ia}
		J.~D.~Barrow,
		Phys. Lett. B \textbf{187}, 12-16 (1987)
		\bibitem{Chakraborty:2023neh}
		M.~Chakraborty, G.~Sardar, A.~Bose and S.~Chakraborty,
		Eur. Phys. J. C \textbf{83}, no.9, 860 (2023)
		\bibitem{Schwarz:2015cma}
		D.~J.~Schwarz, C.~J.~Copi, D.~Huterer and G.~D.~Starkman,
		Class. Quant. Grav. \textbf{33}, no.18, 184001 (2016)
		\bibitem{Vinutha:2023yee}
		T.~Vinutha, K.~Niharika and K.~S.~Kavya,
		Astrophysics \textbf{66}, no.1, 64-83 (2023)
		\bibitem{Leon:2013bra}
		G.~Leon and A.~A.~Roque,
		JCAP \textbf{05}, 032 (2014)
		\bibitem{Datta:2021jwr}
		S.~Datta and S.~Guha,
		Phys. Dark Univ. \textbf{34}, 100890 (2021)
			\bibitem{Kantowski:1966te}
		R.~Kantowski and R.~K.~Sachs,
		J. Math. Phys. \textbf{7}, 443 (1966)
		\bibitem{Leon:2010pu}
		G.~Leon and E.~N.~Saridakis,
		Class. Quant. Grav. \textbf{28}, 065008 (2011)
		\bibitem{Fadragas:2013ina}
		C.~R.~Fadragas, G.~Leon and E.~N.~Saridakis,
		Class. Quant. Grav. \textbf{31}, 075018 (2014)
		\bibitem{Paliathanasis:2022vux}
		A.~Paliathanasis,
		Symmetry \textbf{14}, no.10, 1974 (2022)
		\bibitem{Oliveira-Neto:2021val}
		G.~Oliveira-Neto, D.~L.~Canedo and G.~A.~Monerat,
		Braz. J. Phys. \textbf{52}, no.4, 130 (2022)
		\bibitem{Conradi:1994yy}
		H.~D.~Conradi,
		Class. Quant. Grav. \textbf{12}, 2423-2440 (1995)
		\bibitem{Simeone:2002fp}
		C.~Simeone,
		Gen. Rel. Grav. \textbf{34}, 1887-1893 (2002)
		\bibitem{Barbosa:2004kp}
		G.~D.~Barbosa and N.~Pinto-Neto,
		Phys. Rev. D \textbf{70}, 103512 (2004)
		\bibitem{Collins:1977fg}
		C.~B.~Collins,
		J. Math. Phys. \textbf{18}, 2116 (1977)
	\bibitem{deCesare:2020swb}
	M.~de Cesare, S.~S.~Seahra and E.~Wilson-Ewing,
	JCAP \textbf{07}, 018 (2020)
	\bibitem{Xanthopoulos:1992fh}
	B.~C.~Xanthopoulos and T.~Zannias,
	J. Math. Phys. \textbf{33}, 1420-1430 (1992)
	\bibitem{Mendes:1990eb}
	L.~E.~Mendes and A.~B.~Henriques,
	Phys. Lett. B \textbf{254}, 44-48 (1991)
	\bibitem{vanElst:1995eg}
	H.~van Elst and G.~F.~R.~Ellis,
	Class. Quant. Grav. \textbf{13}, 1099-1128 (1996)
	\bibitem{Marklund:1998sq}
	M.~Marklund and M.~Bradley,
	Class. Quant. Grav. \textbf{16}, 1577-1597 (1999)
	\bibitem{Ghorani:2021xrs}
	E.~Ghorani and Y.~Heydarzade,
	Eur. Phys. J. C \textbf{81}, no.6, 557 (2021)
	\bibitem{Sa:1992aw}
	P.~M.~Sa and A.~B.~Henriques,
	Phys. Lett. B \textbf{287}, 61-68 (1992)
	\bibitem{Bali:2016fdm}
	R.~Bali and S.~Singh,
	Grav. Cosmol. \textbf{22}, no.4, 394-403 (2016)
	\bibitem{Adhav:2008zz}
	K.~S.~Adhav, V.~G.~Mete, A.~S.~Nimkar and A.~M.~Pund,
	Int. J. Theor. Phys. \textbf{47}, 2314-2318 (2008)
	\bibitem{De:2022shr}
	A.~De, S.~Mandal, J.~T.~Beh, T.~H.~Loo and P.~K.~Sahoo,
	Eur. Phys. J. C \textbf{82}, no.1, 72 (2022)
	\bibitem{Koussour:2022wbi}
	M.~Koussour, S.~H.~Shekh and M.~Bennai,
	Phys. Dark Univ. \textbf{36}, 101051 (2022)
	\bibitem{Koussour:2022jss}
	M.~Koussour, S.~H.~Shekh, M.~Govender and M.~Bennai,
	JHEAp \textbf{37}, 15-24 (2023)
	\bibitem{Thorne}
	Thorne, K.S. Primordial Element Formation, Primordial Magnetic Fields, and the Isotropy of the Universe. Astrophysical Journal, 148, 51 (1967).
	\bibitem{Bunn:1996ut}
	E.~F.~Bunn, P.~Ferreira and J.~Silk,
	Phys. Rev. Lett. \textbf{77}, 2883-2886 (1996)
	doi:10.1103/PhysRevLett.77.2883
	[arXiv:astro-ph/9605123 [astro-ph]].
	\end{thebibliography}
\end{document}